\documentclass[12pt]{article}
\usepackage{graphicx}
\usepackage{bm}
\date{}
\begin{document}

\title{{\bf The Thales experiment}}
\maketitle
\author{\hskip 0.5cm Richard KERNER$^*$ } \\

{\small $*$  {Address : 
LPTMC, Universit\'e Paris-VI - CNRS UMR 7600 , \\ Tour 24, 4-\`eme , Boite 121,
4 Place Jussieu, 75005 Paris, France \\ 
Tel.: +33 1 44 27 72 98, \,  Fax: +33 1 44 27 51 00, \\  e-mail : richard.kerner@upmc.fr}}

%
%
%
%
%
%
%
%
%
%
%
%
%
%
%

\vskip 0.5cm
\centerline{\bf 1. $ \;$ Introduction: Thales and the Great Pyramid}
\vskip 0.3cm
\indent
Thales of Miletus ($\Theta \alpha \lambda \eta \sigma, \; \; 625 \sim  547 \; B.C.$) was considered 
by ancient Greeks as one of their {\it Seven Sages},
as the father of scientific approach to the description of natural phenomena. 
and perhaps as the first person deserving the title of mathematician.

Thales became famous for his prediction of solar eclipse of $585 \; B.C.$, and for his ability to evaluate
dimensions of objects at a distance, by comparing their shadows with the shadow of a stick of known dimension.

The relationship of proportionality used by Thales to determine the height 
of the Great Pyramid is also an introduction of {\it linear dependence}, the essence of linear algebra.

It has become such a commonplace, that the physical aspects of this fundamental experience
are rarely considered in a more detailed manner.

In fact, Thales has performed an important physical experiment relating different
definitions of geometry; to put it more precisely, the notions of straight lines and right angles.
It turns out that the phenomena involved in this experiment belong to quite different
domains of physics: gravitation, quantum mechanics and electromagnetism. The fact that  
they lead to three different, but compatible definitions of geometry, suggests that these distinct aspects of physical reality
are apparently related. This fact gives rise to one of the
most important and fundamental questions concerning physics and our perception of physical world, 
still open after more than twenty-five centuries.

\begin{figure}[hbt!]
\centering
\includegraphics[width=3.2cm, height=4cm]{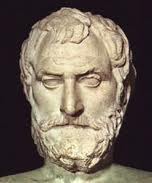}
\hskip 0.5cm
\includegraphics[width=7.7cm, height=4cm]{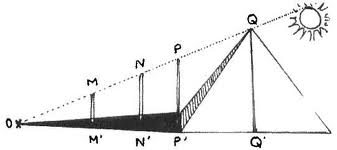}
\caption{\small {Thales of Miletus and the
schematic representation of his experiment: the ratios between  the lengths of vertical objects 
and their shadows are constant,  } {\bf $MM' : OM = QQ' : OQ$ }}
\label{fig:Thalesscheme}
\end{figure}

 
\vskip 0.6cm
\centerline{\bf 2. $\;$ The three definitions of geometry}
\vskip 0.3cm
\indent
Let us analyze the premices and hypotheses that enabled Thales to draw his conclusions
and to state the theorem of parallel lines cutting the angle formed by two intersecting 
straight lines.

The first two assumptions are that the segment { $OQ'$} on the ground is indeed a
straight line, and that the two segments, the height of the pyramid $QQ'$
and the stick $MM'$ are also straight, and form the same angle with the
line of the ground $OM'Q'$  (in this case, the right angle of $90^o$).

This is a physical statement, and the fact that the two objects are straight 
and vertical could be checked using of the well known instruments based on the 
exploitation of gravity.

\begin{figure}[hbt]
\centering
\includegraphics[width=3cm, height=3.2cm]{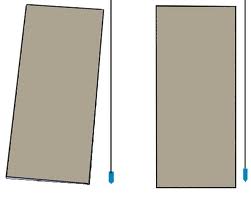}
\hskip 0.5cm
\includegraphics[width=4.1cm, height=2.8cm]{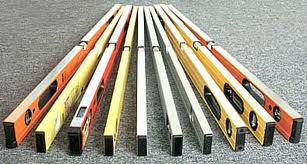}
\caption{\small{Two instruments used for checking whether a straight line or a plane is vertical or horizontal.}}
\label{fig:pionlibella}
\end{figure}

Both instruments shown in Fig. ({\ref{fig:pionlibella}) are based on the use of the gravity field of Earth, 
defining local vertical directions and horizontal
planes (equipotential surfaces).

The fact that the two segments are vertical and straight is based on the assumption that
the string sustaining a heavy object in the gravitational field on the surface
of  Earth may serve as a definition of a {\it vertical straight line}. Checking
the horizontality of the ground is performed using the same principle. To be more precise, the fact that
the string supporting the heavy object takes on the straight shape is due to the tension to which
it is subjected due to the gravitational force. 

A straight line can be obtained in this manner
even inside an artificial satellite orbiting around Earth, in absence of gravitational forces, in total
weightlessness. Any elastic string subjected to tension will take on the form of a straight line.
The tension can be caused by forces having nothing to do with gravitation - e.g. the force of our muscles,
or some mechanical or electrical device.

However, Earth's gravitation is crucial in defining the {\it right angle} between the horizontal ground
surface and the two {\it distant} straight lines, the height of the pyramid  $QQ'$ and the stick $MM'$,
thus determining what is often called {\it distant parallelism}.

Although the Thales theorem seems to concern exclusively spatial relationships between straight lines of certain type,
idealizing spatial interplay between physical objects, {\it time} is implicitly involved in physical hypotheses
necessary to justify the result of Thales' measurements.

When Thales was performing his experiment, the Great Pyramid was more than $2000$ years old, which by the way explains
why its exact dimensions have been since long forgotten. A tacit assumption was that it kept its initial form, including
all angles and dimensions. Even if we exclude the occurrence of seismic events, he had still to admit that the stones forming
the pyramid kept their shape unchanged during very long periods of time.  

The fact that the stick also remains straight and stiff is due to the similar assumption, namely, 
that it is made of a material whose cohesion is sufficient to keep its shape
unchanged (a common definition of a solid body). As seen from our present perspective, this hypothesis is based on the assumption 
that atoms can form stable structures able to keep unchanged under reasonable conditions (e.g. the ambient temperature
not exceeding certain values). From the four-dimensional point of view, this means that atoms and molecules can follow 
parallel timelike geodesics, with null geodesic deviation.
  
Incidentally, the ability of atoms and molecules to form stable periodic structures makes possible an alternative
definition of straight lines and right (and not only right) angles. Crystals represented in Fig. (\ref{fig:wurtzite})
show remarkable linear structure as well as apparently perfect angles, $90^o$ in the case of cubic lattice of $NaCl$,
and $60^o$ and $120^o$ in the case of quartz ($SiO_2$).
\vskip 0.2cm
\begin{figure}[hbt!]
\centering
\includegraphics[width=4.7cm, height=3.6cm]{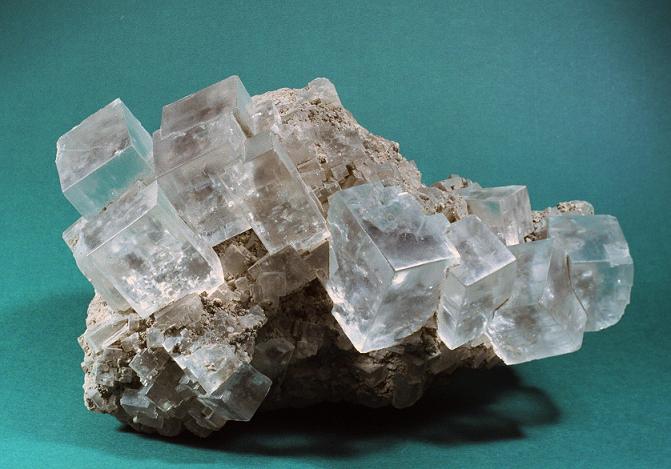}
\hskip 0.3cm
\includegraphics[width=4.5cm, height=3.6cm]{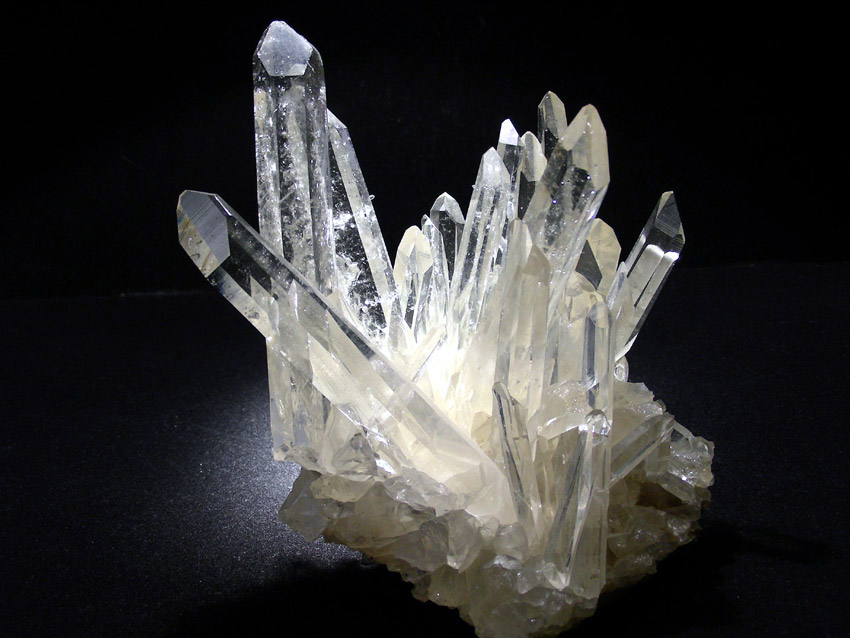}
\hskip 0.3cm
\includegraphics[width=3.3cm, height=3.3cm]{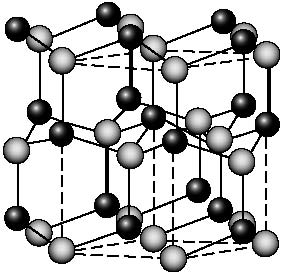}
\caption{\small{Crystals of ordinary salt $NaCl$, of quartz $SiO_2$, and an example of crystalline lattice ($SiO_2$ - wurtzite). 
The interatomic forces impose the shapes and the geometry of solid bodies.}}
\label{fig:wurtzite}
\end{figure}

The straight lines and right angles obtained in the traditional way, by using compass, ruler and a sheet 
of paper, are based on the same physical principle, which is the existence of solid bodies serving as
standards of length. The geometry based on solid bodies' shapes is independent of gravitational field
that determines parallel vertical lines and the horizontal plane in Thales' experiment. From the present
point of view, the existence of stable configurations of atoms, as well as that of atoms themselves, can be understood
only using the principles of quantum mechanics, until now  seemingly independent of gravitational phenomena.

But this is not the end of the story. A third type of straight line is involved in the experiment, the light ray
along the line $QPNMO$. The character of this line is due to the properties of electromagnetic waves' propagation
in vacuo (as far as the influence of air can be neglected), which {\it a priori} is independent of gravitational
phenomena as well as of the forces predominant on the atomic level. Light wavefronts and rays set forth an
alternative notion of straight lines and angles, resulting in {\it conformal geometry}, which preserves the notions 
of straight lines and angles, but ignores the notions of length and distance.\footnote{Thales made 
also an extra tacit assumption, namely, that the properties used for the definition of straight lines, 
parallelism and right angles were scale independent, i.e. they were the same for the small stick and for the
Great Pyramid. The extension towards even greater dimensions, including the Skies, seemed also obvious.}



\vskip 0.6cm
\centerline{\bf 3. $\;$ The three realms of physical world}
\vskip 0.2cm
\indent
The results of Thales' experiment can be interpreted in two ways. In fact, he established 
the coincidence of three completely different definitions of a straight line. The first 
came from the natural shape a string with a heavy body attached to its end takes under the influence of the
gravitational field of the Earth. 

The gravitational field defines also the right angle between the horizontal ground and two distant versicals,
the height of the pyramid and the stick. The mathematical expression of this assertion is given by the
potential function $U(x,y,z)$ defining the {\it equipotential surfaces} $U =  Const$. On the surface of Earth
this equation defines the horizontal plane and the vertical direction, since we have
$$d U = {\bf grad} U \cdot {\rm d} {\bf r} = 0,$$
with the vector
$${\bf grad} U = \left[ \frac{\partial U}{\partial x}, \;   \frac{\partial U}{\partial x}, \;   \frac{\partial U}{\partial x}, \right]$$
defining the local vertical direction, while all displacements orthogonal to it define (locally) a horizontal plane.

The second definition comes from the material shape of the stick. The existence of solid bodies which can be used as standards
of lengths and angles results from symmetry properties of interatomic forces, which in turn can be derived {\it ab initio}
according to the rules of quantum mechanics, valid on the atomic scale.

The third straight line is given by the light ray, which comes from an idealization of electromagnetic wave propagation
from a very distant source. The sunlight illuminating the Earth is well described by a plane wave:
$$A \, \cos (\omega t - {\bf k}\cdot {\bf r}),$$
with planar wavefronts given by the implicit equation $\omega t - {\bf k} \cdot {\bf r} = $ Const. 
The rays are parallel to the wave vector ${\bf k}$, everywhere perpendicular to planar wavefronts.

The first interpretation of the experiment coinciding with what Thales was interested in, is based on the supposition
that all definitions of straight lines and angles do coincide, which enabled him to evaluate the height of the Great Pyramid.

The second interpretation would be, with the height of the pyramid considered as a known quantity, as well as the height of the stick, 
to see the result of Thales experiment as a proof that the light rays follow straight lines
compatible with the two definitions involving gravitation and interatomic forces. Or else, that the straight lines and right angles 
defined by means of gravitational field coincide with those defined by light rays and solid rods.

The three alternative definitions of geometry involved in Thales' experiment are directly related to three
different aspects of our perception of nature.
Since the advent of modern physics, the description of the world surrounding us is based on three essential
realms, already present in the Thales experiment, which are 
\vskip 0.2cm

 $\; \; \; \bullet \; $ {\bf Space and time }

 $\; \; \; \bullet \; $ {\bf Material bodies }

 $\; \; \; \bullet \; $ {\bf Forces acting between them }
\vskip 0.2cm

The three main aspects of our perception of physical reality can be distinctly seen in
the fundamental equation expressing Newton's third law of dynamics:
\begin{equation}
{\bf a} = \frac{1}{m} \; {\bf F}
\label{Newton3}
\end{equation}
shows the relation
between three different realms which are dominant in our perception and description of physical
world: massive bodies (``$m$"), force fields responsible for interactions between the bodies 
(``${\bf F}$") and space-time relations defining the acceleration (``${\bf a}$"). 

The same three ingredients are found in physics of fundamental interactions: we speak of elementary 
particles and fields evolving in space and time we deliberately formulated Newton's law of dynamics
in a slightly unusual way, ${\bf a} = \frac{1}{m} \; {\bf F}$, 
in order to separate the directly observable entity ( ${\bf a}$)
from the product of two entities whose definition is much less direct and clear.

Also, by putting the acceleration alone on the left-hand side, we underline the causal relationship between the
phenomena: the force is the cause of acceleration of mass under its influence, and not vice versa.

{\it In modern language, 
the notion of force is generally replaced by the new concept, the fields of various types.}

The fact that the three ingredients are related by the equation (\ref{Newton3}) may suggest that perhaps 
only two of them are fundamentally independent, the third one being the consequence of the remaining two.

Let us represent the three aspects of theories of fundamental interactions by three orthogonal axes,
as shown in the following figure, which displays also three possible choices of two independent aspects 
of physical reality from which we are supposed to be able to derive the third one. 

\begin{figure}[hbt!]
\centering
\includegraphics[width=4.9cm, height=4.2cm]{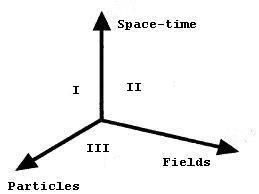}
\caption{\small{The three realms of physics.}}
\label{fig:Threeaxes}
\end{figure}

The attempts to understand physics with only two realms out of three represented 
in (\ref{fig:Threeaxes}) have a very long history. They may be divided in three categories, labeled { $I$, $II$}
 and { $III$}.

In the category $I$ we can easily recognize Newtonian physics, presenting the physical world 
as a collection of material bodies (particles) evolving in absolute space and time, interacting at
a distance. Newton considered light being made of tiny elastic particles obeying the same rules of mechanics
as all material bodies. The notion of fields transmitting forces from one body to another was totally absent.

The controversy concerning the nature of light led to deep differences in the interpretation of space. 
For Huygens, who proved the wave-like propagation of light, space must be filled with some medium enabling the propagation.
Two diametrally opposite views on the status of space and motion prevailed since
then. The Newtonian view was reinforced by Immanuel Kant, who raised the status
of space to the independent and absolute category, existing independently of observers, like the starry sky and the
``moral imperative". (\cite{Kant})

Theories belonging to the category $II$ assume that physical world can be described uniquely as 
a collection of fields evolving in space-time manifold. This approach was advocated by Lord Kelvin, 
A. Einstein, and later on by J.A. Wheeler. The initial impulse was given by 
M. Faraday and J.C. Maxwell, who introduced a revolutionary, anti-Aristotelean and
 anti-Newtonian point of view according to which no interaction at a distance is possible. 
All forces are transmitted by a medium; the space is filled with it. It can be
called ``aether", and the fields of forces become a new physical realm, identified with tensions
inside the aether, which in a sense {\it is} the space. In a sense, ``space" becomes the synonim of
``material continuum", just like from the point of view of a fish, its spatial separation from another fish can be defined 
as the amount of water contained between the two.

As a follower of Maxwell and Faraday, Einstein
believed in the primordial role of fields and tried to derive the equations of motion
as characteristic behavior of singularities of the fields, or of the
space-time curvature. One can say that in Einstein's vision, fields replaced the aether. (\cite{EinsteinInfeld}) 

In the spirit of F. Klein's programme, H. Minkowski defined the hyperbolic geometry of space united with
time in a single entity named {\it ``space-time manifold"}. Its geometry was defined by the action of the
Lorentz-Poincar\'e group. However, at a closer look, physically measurable entities that are subjected 
to Lorentz transformations (the ``four-vectors") are not at all the time and space coordinates, but the
conserved physical quantities, such as energy $E$ and momentum ${\bf p}$,
or the frequency $\omega$ and the wave vector ${\bf k}$ of an electromagnetic wave (or more precisely, of a photon).

The Minkowskian spacetime inherits the Lorentz-Poincar\'e symmetry because it is defined via measurements 
based on photons and their interaction with electrons, whose energy, momentum and spin are
Lorentz-covariant quantities and span representation spaces of the Lorentz-Poincar\'e group.
 
The category $III$ represents an alternative point of view supposing that the existence of matter
is primary with respect to that of the space-time, which becomes an ``emergent" realm - an euphemism
for ``illusion". Such an approach was advocated recently by N. Seiberg and E. Verlinde. (\cite{Verlinde})

 It is true that space-time coordinates cannot be treated on the same
footing as conserved quantities such as energy and momentum; we often forget that they exist
rather as bookkeeping devices, and treating them as real objects is a ``bad habit", as pointed out by
 D. Mermin (\cite{Mermin2009})

Seen under this angle, the idea to derive the geometric 
properties of space-time, and perhaps its very existence, from fundamental symmetries and 
interactions proper to matter's most elementary building blocks seems quite natural.
Many of those properties do not require any mention of space and time 
on the quantum mechanical level, as was demonstrated by M. Born and W. Heisenberg  (\cite{BornJH}, \cite{Dirac}) 
in their version of matrix mechanics, or by J. von Neumann's formulation of quantum theory in terms
of the $C^*$ algebras (\cite{JvNeumann}). The non-commutative geometry is another example of formulation of space-time relationships 
in purely algebraic terms (\cite{MDVRKJM}).

Considering quantum physics as the primary underlying reality of which classical objects are an averaged version,
one is led to conclude that quantum properties of physical objects must be intimately related to the definition of geometry in the first place.
\vskip 0.3cm
\centerline{\bf 4. The Thales experiment from todays' perspective}
\vskip 0.3cm
\indent
Let us come back to the experiment carried out by Thales more than twenty-five centuries ago. According to our analysis,
we can recognize to which physical realm belongs each of three definitions of straight line. The two parallel vertical lines,
the pyramid's height and the stick, are made of wood and stone, which keep their form due to their solid state. Being made of
atoms, the existence and properties of such solids can be derived from rules of quantum mechanics. This is the realm of particles
with mass: nucleons and electrons, which form atoms, then molecules, and finally stable crystalline or amorphous solids. The
electromagnetic forces play also an important role, keeping the electrons around the nuclei, and creating the residual Lennard-Jones
potentials outside the atoms, giving rise to the Van der Waals forces.

The light rays which created also shadows of the pyramid and stick alike are, as we know now, the innumerable swarm of photons
creating a common planar wavefront. They are identified with a massless gauge field, thus belonging to the realm of forces making
possible the interaction between massive charged particles. The interaction between the photons and electrons of atomic outer shells
is described most adequately with the rules of quantum physics. 

Only the third side of each of the two triangles appearing in Thales' experiment, the parallel vertical lines, seem
to have nothing to do with quantum physics, their directions being defined by the gravitational field of Earth. But after
closer scrutiny we can conclude that even in this case the devices made of solids are necessary to detect the presence of
gravitation, and the information about their behavior is carried forth by photons. 

At this point we can ask whether the Thales experiment could be performed without gravity - and the answer is "yes". 
To construct a plane and two vertical parallel lines the solid standards of length and right angle would suffice,
it can be done with standard compass and ruler. Therefore, the experience following Thales' scheme, can be viewed
as checking whether the laws of gravity are compatible with the geometry defined by solid bodies and light, i.e.
by the classical limit of quantum mechanics and quantum field theory. By the way, with very precise measurements
of angles we would be able to find out the actual curvature of Earth surface, because the verticals defined by its
gravitational field are not parallel in fact: the distance of about $31$ metres corresponds to one second of arc
between the local vertical directions defined by Earth's gravitation. 

The present analysis of Thales' experiment suggests that among the three realms of physics represented in (\ref{fig:Threeaxes}),
particles and fields (quantum physics) define the geometry when they constitute classical objects like solid bodies
and wavefronts, while the presence or absence of gravitation is checked with the help of other classical objects.
To put it in a very rough manner, solid bodies made of atoms and wavefronts made of photons are there no matter whether 
gravity exists or not; on the contrary, gravity, as well as the geometry of space-time itself, is defined through the properties
of solid bodies and light rays. The very detection of gravitational effects cannot be performed without extended massive bodies, 
behaving like classical objects. Even the famous experiment confirming the variation of proper time under the influence of gravity, 
performed by Pound and Rebka (\cite{PREB}) in 1959, uses the M\"ossbauer effect based on the collective behavior of crystalline lattice 
which cancels the recoil effect during photon absorption.

Thales' theorem led the way to all subsequent measurements of great distances, first on land and sea, then applied to the measurement
of radius of the Earth by Eratosthenes, then for determining astronomical distances by Aristarchos of Samos. Later on
the measurement of distance to the closest stars due to the observed annual parallax is just another application of the
Thales theorem. The determination of shapes of planetary orbits by Kepler was based on triangulation, which is also
a variety of the same theorem. The subsequent determination of the true dimensions of Solar System was made only
in 1769 due to the observation of Venus' transit and the exact knowledge of longitude by Captain Cook who performed
the observations on the island of Tahiti. The crucial measurement concerned the exact time of the phenomenon as observed
from distant places on Earth. The longitude could be determined also due to the invention of chronometre by Huyghens.

The speed of light in the vacuum being constant for all Galilean observers, nowadays the measurements of
distances in space can be replaced by precise measurements of time delays, like with the Global Positioning System (GPS). And it is not accidental
that very large distances are measured in time equivalents, the light-years. The exact measurements of time, which nowadays
attains the precision of $10^{-12}$ second, enable us to determine distances with similar degree of precision - less than $1$ cm
on the surface of our globe. Such time measurements are possible due to atomic clocks obeying quantum mechanical rules.

All the information we receive from the surrounding world is carried by photons, leptons and baryons, elementary particles
whose properties and behavior are extremely well described by quantum physics. However, we can perceive and analyze them
only through devices representing the classical limit of quantum mechanics. No wonder that the geometry built on the base of
the obtained data reflects the symmetry group acting in the space of states of elementary particles - the Lorentz-Poincar\'e
group. The transformation properties of conserved physical entities such as the four-vectors $k^{\mu} = [ \frac{\omega}{c}, {\bf k}]$
or $P^{\mu} = [ \frac{E}{c}, {\bf p} ]$ are extended to the {\it dual space} of differential forms $d x_{\mu} = [c dt, d {\bf x}]$.
These, in turn, are defined experimentally using classical objects, whose very existence (rigid bodies made up from atoms,
light wavefronts made out of photons) is explained by quantum theory.

Thus the conclusion in the case of Thales experiment is that in order to construct the Euclidean geometry of space, only these
two physical phenomena were needed, the light playing the role of the ruler (defining the straight lines), and the rigid
bodies playing the role of compass (defining distances). Gravity was used to define parallel straight lines and right angles,
but its use was not necessary. On the contrary, its influence can be measured using exclusively classical objects.  

Apparently, the gravitation can be perceived only in the classical limit, and not on the quantum level. In spite of
 numerous attempts, there is no quantum limit of classical physics. This suggests two conclusions:
\vskip 0.3cm
 first, that space and its geometry are defined only in the classical limit of quantum theory;
\vskip 0.2cm
second, that gravity is also a classical phenomenon, appearing only when the collective effects can be perceived, just like
classical thermodynamics can be defined only as a limit of statistical physics
\vskip 0.3cm
In this case, quantizing gravitational waves is
as hasardous an enterprise as an attempt to quantize the waves on the surface of water.

\end{document}